\documentclass[10pt,fleqn]{article}
            \newcommand{\be}{\begin{eqnarray}}
            \newcommand{\ee}{\end{eqnarray}}
\newcommand{\e}[1]{\label{e:#1}\end{eqnarray}}
     \newcommand{\eg}{{\em e.g.\ }}

            \newcommand{\del}{{\delta}}

           \newcommand{\ra}{{\rightarrow}}
 \newcommand{\lora}{{\longrightarrow}}

            \newcommand{\beq}{\begin{quote}}
            \newcommand{\eq}{\end{quote}}
            
            \newcommand{\al}{\alpha}
            \newcommand{\ben}{\begin{enumerate}}
            \newcommand{\een}{\end{enumerate}}
            \newcommand{\bit}{\begin{itemize}}
            \newcommand{\ei}{\end{itemize}}
    	
            \newcommand{\re}[1]{(\ref{e:#1})}
            \newcommand{\edfl}[1]{\label{#1}\end{df}}
\newcommand{\vb}{{\cal h}}
\newcommand{\hb}{{\cal i}}

\newcommand{\te}{{\theta}}

	\newcommand{\halv}{{\frac{1}{2}}}
	\newcommand{\kvart}{{\frac{1}{4}}}

\def\cC{{\cal C}}
  \def\bcC{\bar{{\cal C}}}
  \def\cP{{\cal P}}
 \def\cH{{\cal H}}
  \def\cL{{\cal L}}
  \def\bcP{\bar{{\cal P}}}

\begin{document}
\noindent
G\"{o}teborg ITP 97-10\\
May 1997\\
\begin{center}{\Large\bfseries Time evolution in general gauge
theories}\footnote{Talk at the International Workshop ``New Non 
Perturbative Methods and Quantization on
       the Light Cone", Les Houches, France, Feb.24-March 7, 1997}
\end{center}
\vspace*{3 mm}
\begin{center} 
\begin{center}Robert
Marnelius\footnote{E-mail: tferm@fy.chalmers.se}\\ \vspace{2mm}
{\it
Institute of Theoretical Physics, Chalmers University of Technology,\\
G\"{o}teborg University,
S-412 96  G\"{o}teborg, Sweden}\end{center}
\vspace*{5 mm}
\end{center}
\section*{Introduction}
In this talk I will discuss some properties of time evolutions in general gauge
theories within a BRST quantization \cite{Time}. More precisely I will discuss the
choices of Hamiltonians within the Hamiltonian  framework set up by Batalin, Fradkin
and Vilkovisky which is called the BFV formulation \cite{BFV} (for a review see \eg
\cite{BF}). This I will do from the point of view of
 an operator formulation for inner product solutions within the BFV
scheme which I have been developing during some years  \cite{Simple}-\cite{Proper}.
This formalism turns out to yield  more information about quantum properties
than just  an effective BRST invariant Lagrangian or Hamiltonian
formulation.  In fact, an effective Hamiltonian
is more difficult to extract within this scheme, but the procedure
provides for a deeper understanding of  the
standard BFV prescriptions. These results will be briefly reviewed. As a particular
example of a natural consequence I will at the end show that QED is coBRST invariant.
However, let me first review the standard BFV formulation.
\section*{Standard BFV-BRST}
Within the BFV formulation Hamiltonians of general 
gauge theories are assumed to have the form
\be
&&H=H_0+\int v_i\te_i,
\e{1}
where $v_i$ are Lagrange multipliers and $\te_i$ 
constraint variables. (Repeated indices are summed over and integrals are over 
space coordinates.) $H_0$ and
$\te_i$ satisfy the super Poisson bracket conditions
\be
&&\{H_0, \te_i\}=C_{ij}\te_j,\;\;\;\{\te_i, \te_j\}=C_{ijk}\te_k.
\e{2}
where $C_{ij}$ and $C_{ijk}$ may be functions on the phase space. 
In the corresponding BRST quantization BFV 
introduces the following additional degrees of freedom:
\bit
\item{$\pi_i$ -- conjugate momenta to the Lagrange multipliers $v_i$. 
(They are additional abelian
constraint variables.)}
\item{$\cC_i$, $\cP_i$ -- ghosts  and their conjugate momenta.}
\item{$\bcC_i$, $\bcP_i$ -- antighosts  and their conjugate momenta.}
\ei
Their Grassmann parities and ghost numbers are
\be
&&\varepsilon(\cC_i)=\varepsilon(\bcC_i)=\varepsilon(\te_i)+1,
\;\;\;gh(\cC_i)=1=-gh(\bcC_i).
\e{3}
In this extended phase space the BRST charge is given by
\be
&&Q=\int\left(\cC_i\te_i+\ldots+\bcP_i\pi_i\right),
\e{4}
where the dots indicates terms determined by the super Poisson 
bracket condition $\{Q, Q\}~=~0$.
The Hamiltonian for the effective BRST invariant theory is defined to be
\be
&&H_{BRST}=H'_0+i\{Q, \psi\},\;\;\;\{H_{BRST}, Q\}=0,
\e{5}
where
\be
&&H'_0=H_0+\mbox{ghost dependent terms},\;\;\;\{H'_0, Q\}=0,
\e{6}
and where in turn $\psi$ is an odd gauge fixing fermion 
which usually is chosen such that
\be
&&H_{BRST}=H+\int(\;)\pi_i+\mbox{ghost dependent terms}.
\e{7}
Such a Hamiltonian leads to a BRST invariant effective 
Lagrangian of the standard form
\be
&&\cL_{BRST}=\cL+\cL_{gf}+\cL_{gh}.
\e{8}
The general allowed form for $\psi$ as prescribed by BFV is
\be
&&\psi=\int\left(\cP_iv_i+\bcC_i\chi_i\right),
\e{9}
where $\chi_i$ are gauge fixing variables to $\te_i$. 
(The matrix $\{\chi_i, \te_j\}$ is required
to be invertible.)

One may observe that neither $H'_0$ nor $\psi$ are 
uniquely determined. For instance, in an
abelian gauge theory $H_{BRST}$ is invariant under the transformations
\be
&&H_0\,\lora \,H_0+x_i\te_i,\;\;\;\psi\,\lora \,\psi+\cP_ix_i
\e{10}
for any BRST invariant variable $x_i$.
\section*{Operator quantization on inner product spaces.}
\subsection*{Case 1: $H'_0=0$}
This case includes all reparametrization invariant 
theories, such as particles, strings, and
gravity. The operator quantization proceeds here as 
follows: Quantize all degrees of freedom and
construct an extended inner product state space $V$. 
Physics is then what is contained in the
subspace
$V_{ph}\subset V$ defined by $QV_{ph}=0$. $V_{ph}$ is 
degenerate since the zero norm states $QV$ is
contained in $V_{ph}$. The nondegenerate inner 
product space is therefore $V_s=V_{ph}/QV$, the
states of BRST singlets. $V_s$ is an inner product 
space if $V$ is an inner product space. An
important concept in this connection is the coBRST 
charge $^*Q$ \cite{CoB}. It is defined by
\be
&&^*Q=\eta Q\eta,
\e{11}
where $\eta$ is an hermitian metric operator such 
that $\eta^2=1$ and $\vb u|\eta|u\hb\geq0$
$\forall |u\hb\in V$. Thus, $\eta$ maps $V$ onto a Hilbert
 space and $^*Q$ is just the hermitian
conjugate of $Q$ in this Hilbert space. We have $^*Q^2=0$.
 In terms of the coBRST charge the BRST
singlets $|s\hb\in V_s$ are determined by
\be
&&Q|s\hb=\,^*Q|s\hb=0
\e{12}
or equivalently
\be
&&\triangle|s\hb=0,\;\;\;\triangle\equiv [Q,\,^*Q]_+.
\e{13}
Now it is usually very difficult to find the appropriate inner product space
$V$. (There are even cases which allow for several 
different choices.) Fortunately, there is a possibility to
construct formal operator expressions for the singlets 
$|s\hb$ without prescribing $V$. In fact,
such expressions will at the end tell you the appropriate 
prescription for $V$ \cite{Proper}. Since this
formalism is not yet completely rigorously proved, I will
 present the main ingredients as
a set of proposals:
\beq
{\sl Proposal 1}: If $Q=\del+\del^{\dag}$, where 
$\del$, $\del^{\dag}$ are independent nilpotent
operators each containing effectively  half the 
constraints of $Q$, then the solutions of
$\del|ph\hb=\del^{\dag}|ph\hb=0$ are formally 
inner product solutions what concerns the unphysical
degrees of freedom.\eq
\beq
{\sl Proposal 2}: $Q$ for any gauge theory in BFV 
form may be decomposed as $Q=\del+\del^{\dag}$,
where
$\del$,
$\del^{\dag}$ are independent and each containing 
effectively half the constraints of $Q$ and
such that
$\del^2=0$ and $[\del,
\del^{\dag}]_+=0$.\eq
\beq
{\sl Proposal 3}: The formal solutions of
$\del|ph\hb=\del^{\dag}|ph\hb=0$ have up to zero norm 
states the general form $|ph\hb=e^{[Q,
\psi]_+}|\phi\hb$ where $\psi$ is an odd gauge fixing 
fermion of the form \re{9}, and where
$|\phi\hb$ satisfies simple hermitian conditions.\eq 
Proposal 1 has been shown to be valid in all
investigated cases. Proposal 2 has been proved for 
general Lie group theories \cite{Simple}.
Concerning proposal 3 the following may be said: Formal 
inner product solutions of the
form
$|ph\hb=e^{[Q,
\psi]_+}|\phi\hb$ exist for any gauge theory if
$Q$ is in BFV form. In fact, in \cite{BRM} it was shown that 
the BRST singlets $|s\hb$ locally may
be written as $|s\hb=e^{[Q,
\psi]_+}|\phi_s\hb$ where $|\phi_s\hb$ is a ghost and gauge 
fixed $|\phi\hb$. There it was shown that
$D_i|s\hb=0$ where $D_i$ are a complete set of BRST doublets 
($C$,$[Q, C]_\pm$-pairs) satisfying $[D_i,
D_j]_\pm=c_{ijk}D_k$, and that $[D_i, D_j^{\dag}]_\pm$ is an 
invertible matrix operator. ($\vb\phi|\phi\hb$
is undefined while $\vb\phi|e^{[Q, \psi]_+}|\phi\hb$ is well 
defined and independent of $\psi$.)

{ What is the form of the coBRST charge within the BFV formalism?}
It turns out that the coBRST charge has the form of an allowed gauge fixing
fermion~\cite{GRM}. However, one may notice that
$|s\hb=e^{[Q,\,^*Q]_+}|\phi_s\hb=e^\triangle|\phi_s\hb$ does not satisfy
$^*Q|s\hb=0$. Only
$|s'\hb=U|s\hb$ does, where $U$ is a unitary operator\cite{GRM}.\\ \\
\subsection*{Case 2: $H'_0\neq0$}
This case may always be transformed to case 1 ($H'_0=0$) by making the gauge theory
reparametrization invariant. The transformed theory may then be treated as before.
This is the fundamental approach here.
\emph{Thus, nontrivial time evolution does not cause any basic problems.} However, 
problems do occur
 whenever one   tries to find  inner product solutions without going to the
corresponding reparametrization invariant theory. A simple and natural
prescription  for inner product solutions in the latter case is first to
construct inner product solutions as in case 1 ignoring $H'_0$ and then to
require 
$|ph, t\hb$ or $|s, t\hb$ to be determined by a Schr\"odinger equation with the
Hamiltonian $H'_0$. From the corresponding reparametrization invariant theory one
finds that this is possible  provided the
gauge fixing conditions satisfy some weak conditions like
\be
&&[Q, [H'_0,\psi]_-]_+=0,\;\;\;[\psi, [H'_0,\psi]_-]_+=0.
\e{14}
That the Hamiltonian must be $H'_0$ follows from the fact that 
the BRST charge in the corresponding
reparametrization invariant theory is
\be
&&\tilde{Q}=Q+\cC(\pi+H'_0)+\bcP\pi_v,
\e{15}
where $\cC$ and $\bcP$ are new ghost variables, $\pi_v$ 
is the conjugate momentum to a new Lagrange
multiplier, and $\pi$ is the conjugate momentum to a dynamical 
time variable. Thus, since $\pi+H'_0$ is a
new constraint variable it is easily understood that 
$(\pi+H'_0)|\;\hb=0$ is a natural equation, and
 this is the Schr\"odinger equation with $H'_0$ as Hamiltonian. 
It turns out that the BRST singlets
satisfy this Schr\"odinger equation strictly 
under weak conditions like \re{14}. However, the problem with this procedure is that
$\psi$ and $H'_0$ are not uniquely given. It is well known that one by unitary
transformations may change the constraint variables in the BRST charge and $H_0'$ is
part of a constraint in the reparametrization invariant theory (15). On the other
hand, this freedom may be used to find a $H'_0$ and $\psi$'s satisfying conditions
(14). Whether or not this is possible in general is unclear though.

According to the procedure above we have in the case when $H'_0$ has no explicit time
dependence 
\be
&&|ph, t\hb=e^{-iH'_0t}|ph\hb.
\e{16}
Since the first condition in \re{14} implies 
$[H'_0, [Q, \psi]_+]_-=0$ by means of the Jacobi
identities, \re{16} combined with proposal 3 implies
\be
&&|ph, t\hb=e^{-iH'_0t+[Q, \psi]_+}|\phi\hb.
\e{17}
This leads to
\be
&&\vb ph, t'|ph, t\hb=\int d\omega' d\omega\, {\phi'}^*({\omega'}^*)
\phi(\omega)\vb \omega', t'|\omega^*, t \hb,
\e{18}
where $\omega$ denotes all coordinates of the original BRST 
invariant theory. (Due to the indefinite metric state space not all hermitian
operators have real eigenvalues.) After the replacement
$\psi\ra(t'-t)\psi/2$, which is possible for any finite $t'-t\neq0$, 
one may
derive the path integral representation
\cite{Path,Time}:
\be
&&\vb \omega', t'|\omega^*, t\hb
=\int D\omega D\pi_\omega\:
\exp{\left\{i\int_t^{t'}\left(\pi_\omega\dot{\omega}-H_{BRST}\right)\right\}},
\e{19}
where 
\be
&&H_{BRST}=H'_0+i\{Q,
\psi\}
\e{20}
is the effective Hamiltonian function in agreement with the BFV prescription (5).
Notice, however, that $H_{BRST}$ is not real in general. Often a real effective
Hamiltonian requires us to choose an imaginary time
$t$
\cite{Path}.

One may notice that the second condition in \re{14} 
is satisfied if $\psi^2=0$. If $[H'_0, \psi]_-=0$
and $\psi^2=0$ then the effective Hamiltonian $H_{BRST}$ 
in the path integral \re{19} satisfies
$\{H_{BRST}, \psi\}=0$ which implies that $\psi$ generates 
a new nilpotent symmetry. That this may
be realized is shown in the following example.

\section*{Example: QED}
Consider for simplicity the Lagrangian density for a 
free electromagnetic field (the metric
is time-like)
\be
&&\cL=-\kvart F^{\mu\nu}F_{\mu\nu},\;\;\;F_{\mu\nu}\equiv 
\partial_\mu A_\nu-\partial_\nu A_\mu.
\e{21}
The canonical momenta to $A_\mu$ are
\be
&&E^\mu={\partial\cL\over\partial\dot{A}_\mu}=F^{\mu 0}.
\e{22}
$E^0=0$
 is a primary constraint. The Hamiltonian density, which is equal to the
canonical energy density $T^{00}$, is given by
\be
&&\cH=-\halv E^iE_i+E^i\partial_i A^0+\kvart F^{ij}F_{ij}.
\e{23}
The Hamiltonian equations of motion are generated by
\be
&&H\equiv\int d^3x \left(\cH(x)+\dot{A^0}E^0\right),
\e{24} 
where 
$\dot{A^0}$  is an arbitrary function which represents the gauge freedom.
(The Lorentz condition
$\partial_\mu A^\mu=0$  demands $\dot{A^0}=-\partial_iA^i$.)
Since
\be
&&\dot{E}^0(x)=\{{E}^0(x), H\}=\partial_iE^i(x)
\e{25}
consistency requires the secondary constraint (Gauss' law) $\partial_iE^i=0$.

The standard Faddeev-Popov Lagrangian for QED is
\be
&&\cL=-\kvart F^{\mu\nu}F_{\mu\nu}-{1\over 2\al}(\partial_\mu
A^\mu)^2-i\partial_\mu\bcC\partial^\mu \cC,
\e{27}
where $\al$ is a real parameter. The corresponding 
Hamiltonian within the BFV scheme is given by
\re{20} where  the standard choice is
\be
&&\cH_0=-\halv E^iE_i+\kvart
F^{ij}F_{ij},\;\;\;Q=\int d^3x(\cC\partial_iE^i-\bcP E^0).
\e{28}
(We have a minus sign in $Q$ since $\pi_v=-E^0$.) The gauge fixing fermion is
\be
&&\psi=\int d^3x(\cP v+\bcC\chi),
\e{29}
where $v=-A^0$ and $\chi=\partial_iA^i+ E^0/{2\al}$. However, this $\psi$ is neither
conserved nor nilpotent, and it does not satisfy the conditions \re{14}.

Now there is another option for $\cH_0$, namely
\be
&&\cH_0=-\halv E^iE_i+{1\over 2\nabla^2}(\partial_i E^i)^2+\kvart
F^{ij}F_{ij},\;\;\;\nabla^2\equiv -\partial_i\partial^i.
\e{30}
This choice determines  the Lagrange multiplier $v$ to be
\be
&&v\equiv-A^0-{1\over 2\nabla^2}\partial_iE^i.
\e{31}
Exactly the same effective Hamiltonian as before is now 
obtained by the formula \re{20} with
$\cH_0$ given by \re{30} and $\psi$ given by \re{29} now with $v$ as in \re{31} and
$\chi=\partial_iA^i+ E^0/{2\al}$. In distinction to the previous construction 
$\cH_0$ in
\re{30} satisfies the strong condition
$\{\psi, H_{0}\}=0$. For $\al=1$ (the Feynman gauge) 
$\psi$ is furthermore nilpotent and may be
identified with a coBRST charge. In this case we have 
$\{\psi, H_{BRST}\}=0$ and $\psi$ generates a symmetry transformation. It is
\be
&&rA^0=\halv\bcC,\;\;\;rA^i=-{1\over 2\nabla^2}\partial^i\dot{\bcC},
\;\;\;r\cC=\halv
i(A^0-{1\over\nabla^2}\partial_i\dot{A}^i),\;\;\;r\bcC=0.
\e{32}
(Like the BRST transformation
it is only nilpotent on-shell.) Of course, also the bosonic charge, $i\{Q, \psi\}$,
is conserved and generates a symmetry transformation. 
The symmetry transformation \re{32} was also given
in
\cite{MLM}. (I am thankful to Joaquim Gomis for pointing out this reference to me.)

 The
corresponding construction for Yang-Mills theories is more difficult. 
The standard BRST fixed
Lagrangian does not allow for a conserved coBRST charge 
due to the Gribov ambiguities \cite{Gri}. ($\chi$ contains the Coulomb gauge
$\partial_iA^i_a$.)


\begin{thebibliography}{Simple}

\bibitem{Time}R. Marnelius, 
 \ {\em  Time evolution in general gauge theories on inner product spaces},
 \ {\sl Nucl.
Phys.}\ {\bf B} (in press)

\bibitem{BFV}I. A. Batalin and G. A Vilkovisky, \ {\sl  Phys. Lett.}
\ {\bf B69}, 309 (1977)\\
 E. S. Fradkin T. E. Fradkina, \ {\sl  Phys. Lett.}
\ {\bf B72}, 343 (1978)\\
I. A. Batalin and E. S. Fradkin, \ {\sl  Phys. Lett.}
\ {\bf B122}, 157 (1983)

\bibitem{BF}I. A. Batalin and E. S. Fradkin, \ {\sl Riv. Nuovo Cim.} \
{\bf 9}, 1 (1986)


\bibitem{Simple}R. Marnelius, \ {\sl Nucl.
Phys.}\ {\bf B395}, 647 (1993);\\
{\sl Nucl.
Phys.}\ {\bf B412}, 817 (1994)

\bibitem{Path}R. Marnelius, 
 \ {\sl  Phys. Lett.}\ {\bf B318}, 92 (1993)

\bibitem{BRM}I. Batalin and R. Marnelius, \ {\sl Nucl.
Phys.}\ {\bf B442}, 669 (1995)

\bibitem{GRM}G. F\"ul\"op and R. Marnelius, \ {\sl Nucl.
Phys.}\ {\bf B456}, 442 (1995)

\bibitem{Proper}R. Marnelius, 
 \ {\sl Nucl. Phys.}\ {\bf B418}, 353 (1994)

\bibitem{CoB}
 K. Nishijima, \  {\sl Nucl. Phys.}\ {\bf
B238}, 601 (1984)\\
M. Spiegelglas, \ {\sl Nucl. Phys.}\ {\bf B283}, 205 (1987)\\
 W. Kalau, J.W. van Holten ,  \ {\sl Nucl. Phys.}\ {\bf B361},  233 (1991)


\bibitem{MLM}M. Lavelle and D. McMullan, 
 \ {\sl Phys.Rev.Lett.}\ {\bf 71}, 3758 (1993)

\bibitem{Gri}V. N. Gribov, \  {\sl Nucl. Phys.}\ {\bf
B139}, 1 (1978) 

\end{thebibliography}
\end{document}